%% file: Frascati_2007_proc.tex
\documentclass{PoS}

\usepackage[latin1]{inputenc}
\usepackage{amsbsy,amsmath,amsfonts,amssymb}
\usepackage{capt-of}

\input{pretext}

\title{Lepton Universality Tests with Kaons}

\ShortTitle{Lepton Universality Tests with Kaons}

\author{\speaker{Rainer Wanke}%
         \thanks{On behalf of the Flavianet Working Group on Kaon Decays
                 ({\tt http://www.lnf.infn.it/wg/vus/})}\\       
        Institut f\"ur Physik, Universit\"at Mainz, D-55099 Mainz, Germany\\
        E-mail: \email{Rainer.Wanke@uni-mainz.de}}

\abstract{Precision data on $K_{l3}$ and $K_{l2}$ decay rates 
          and form factors allow us to perform significant tests of 
          lepton universality and to constrain the strength of non-standard 
          interactions. The present status of these tests 
          and new physics searches are discussed, as obtained by combining all the available 
          results of the various kaon physics experiments.}

\FullConference{KAON International Conference\\
	        May 21-25 2007\\
		Laboratori Nazionali di Frascati dell'INFN, Rome, Italy}

\begin{document}

\section{Introduction}

While lepton flavour violation has recently been discovered in the neutrino sector,
lepton universality in meson decays is strictly required in the Standard Model (SM).
Violation of lepton universality would be an immediate indication of new physics beyond the SM,
and indeed, most theories of physics beyond the SM --- as e.g.\ Supersymmetry --- 
predict lepton flavour violating transitions.

There are mainly two decay modes in kaon physics, in which lepton universality can be 
investigated: the semileptonic decays $K_{l3}$ are sensitive to lepton flavour violation 
in the vector current, while the leptonic decays $K_{l2}$ test the axial current.
So far, precision on lepton universality tests have been poor in the kaon system, 
compared to e.g.\ tests in pion decays.
With new data, which have been taken and analyzed recently by many experiments, 
the precision of these measurements has improved significantly;
kaon decays are now more than competitive with lepton flavour violation searches in $\tau$ and $\pi$ decays.

This report reviews the recent measurements of lepton flavour violation and new physics searches, 
as well as the planned improvements in $K_{l2}$ decays.

\section{Tests of Lepton Universality in $K_{l3}$ Decays}

Search for lepton flavour violation (LFV) in the semileptonic decays $K_{e3}$ and $K_{\mu3}$
is a test of the vector current of the weak interaction. It can therefore be compared
to LFV tests in $\tau$ decays, but is different to LFV searches in
$\pi_{l2}$ and $K_{l2}$ decays.
Currently, the best test of lepton universality in the vector current comes from the comparison
of the $\tau \to e \bar{\nu}_e \nu_\tau$ and $\tau \to \mu \bar{\nu}_\mu \nu_\tau$ decay rates~\cite{bib:pdg},
from which the ratio of coupling constants can be determined to $g_\mu^2/g_e^2 = 0.9998 \pm 0.0040$.
It agrees perfectly with unity within a precision of $0.4\%$.

The $K_{l3}$ decay rate, including possible photons from internal bremsstrahlung, is given by~\cite{bib:leutwylerroos}
\be
\Gamma(K_{l3(\gamma)}) = \frac{G_F^2 m_K^5}{192 \pi^3} C_K^2 S_{EW} |V_{us}|^2 |f_+(0)|^2 I_K^l (1+\delta_K^l)^2.
\ee
Here, $f_+(0)$ is the form factor at $q^2 = 0$, $S_{EW} = 1.0232$ is a short distance
electro-weak correction, $I^l_K$ is the phase space integral which depends on the form factors,
and $V_{us}$ the CKM matrix element.
The factor $C_K^2$ is 1 for $K^0$ and $\frac{1}{2}$ for $K^\pm$. 
The correction $(1+\delta_K^l)^2 \approx 1 + 2 \delta^l_{SU(2)} + 2 \delta^l_{EM}$
takes into account $SU(2)$ symmetry breaking ($\delta^l_{SU(2)}$) and long distance
electro-magnetic interactions ($\delta^l_{EM}$).

In the ratio $\Gmue$, the lepton flavour independent factors 
$C_K$, $S_{EW}$, $|V_{us}|$, $|f_+(0)|$, and $\delta^l_{SU(2)}$ cancel and need not to be considered.
The electro-magnetic corrections have been computed on one-loop level~\cite{bib:cirigliano} and are on the order of 
a few per mil (Tab.~\ref{tab:kl3radcorr}).
Possible correlations between the corrections for the different decays are not taken into account.
\begin{table}[t]
  \begin{center}
      \begin{tabular}{cccc}
      \hline \hline
	            & $\delta^e_{EM}$ $[\%]$ & $\delta^\mu_{EM}$ $[\%]$ & $(1 + \delta_K^\mu)^2/(1 + \delta_K^e)^2$ \\ \hline
       $K^0_{l3}$   & $+0.52(10)$            & $+0.80(15)$              & $1.006(4)$    \\
       $K^\pm_{l3}$ & $+0.03(10)$            & $-0.12(15)$              & $0.997(4)$    \\     
      \hline \hline
      \end{tabular}
     \caption{Radiative corrections to the $K_{l3}$ decay rate~\cite{bib:cirigliano}.}
     \label{tab:kl3radcorr}
  \end{center}
\end{table}
For the phase space integrals, 
the form factor values from the global fit of the Flavianet working group on kaon decays are taken, 
assuming lepton universality for the $K_{e3}$ and $K_{\mu3}$ form factors. 
Using a quadratic expansion for the vector form factor, and a linear for the scalar form factor, 
the fit results on the slopes are $\lambda_+' = (24.82 \pm 1.10) \times 10^{-3}$, 
$\lambda_+'' = (1.64 \pm 0.44) \times 10^{-3}$, and $\lambda_0 = (13.38 \pm 1.19) \times 10^{-3}$~\cite{bib:palutan},
which lead to the phase space integrals given in Tab.~\ref{tab:kl3phasespace}.
\begin{table}[t]
  \begin{center}
      \begin{tabular}{cccc}
      \hline \hline
	            & $I_K^e$       & $I_K^\mu$      & $I_K^\mu/I_K^e$ \\ \hline
       $K_{L,l3}$   & $0.15454(29)$ & $0.10209(31)$  & $0.6617(16)$    \\
       $K^\pm_{l3}$ & $0.15889(30)$ & $0.10504(32)$  & $0.6611(16)$    \\    
      \hline \hline
      \end{tabular}
     \caption{Phase space integrals for $K_{l3}$~\cite{bib:palutan}.}
     \label{tab:kl3phasespace}
  \end{center}
\end{table}

This results to a Standard Model expectation of
\be
R_{K_{\mu3}/K_{e3}}^\text{SM} \equiv \frac{\Gamma(K_{\mu3})}{\Gamma(K_{e3})} = 
                                                               \frac{I_K^\mu}{I_K^e} \, \frac{(1 + \delta_K^\mu)^2}{(1 + \delta_K^e)^2} 
                = \left\{ \begin{array}{ll} \! 0.6657(31) &   \text{for} \: K_{L}.  \\*[1mm]
                                            \! 0.6591(31) &   \text{for} \: K^\pm.  \\
                         \end{array}
                  \right.
\ee

Apart from experiments of the 1970's, five recent and more precise direct measurements~\cite{bib:KLe3_KTeV,bib:KLe3_KLOE,bib:Ke3_E246,bib:Ke3_NA48,bib:Ke3_KLOE} of $R_{K\mu3/Ke3}$ exist
(see Tab.~\ref{tab:kmu3ke3} and Fig.~\ref{fig:kmu3ke3}). All of them are in agreement with the SM expectation.

\begin{table}[t]
  \begin{center}
    \begin{minipage}[t]{0.54\linewidth}
    \begin{tabular}{lcc}
                                           & \\*[-7mm]
      \hline \hline
       Experiment                          & Channel & $\Gamma(K_{\mu3})/\Gamma(K_{e3})$  \\ \hline
       KTeV (2004)~\cite{bib:KLe3_KTeV}    & $K_L$   & $0.6640 \pm 0.0026$   \\
       KLOE (2006)~\cite{bib:KLe3_KLOE}    & $K_L$   & $0.6734 \pm 0.0059$   \\ \hline
       KEK-E246 (2001)~\cite{bib:Ke3_E246} & $K^+$   & $0.671 \pm 0.011$     \\
       NA48/2 (2007)~\cite{bib:Ke3_NA48}   & $K^\pm$ & $0.663 \pm 0.003$     \\
       KLOE (2007)~\cite{bib:Ke3_KLOE}     & $K^\pm$ & $0.6511 \pm 0.0087$   \\
      \hline \hline
                                           & \\*[-1mm]
    \end{tabular}
    \caption{Recent direct measurements of $\Gmue$.}
    \label{tab:kmu3ke3}
    \end{minipage}
    \hspace*{\fill}
    \begin{minipage}[t]{0.41\linewidth}
      \raisebox{-17mm}{\includegraphics[width=\linewidth]{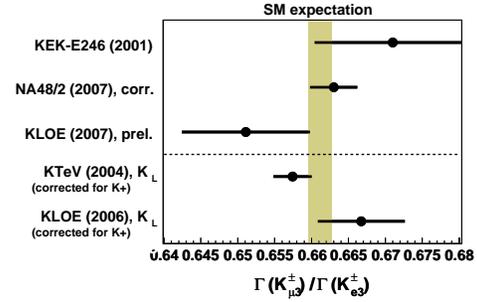}}
      \captionof{figure}{Measurements and expectation for $\Gamma(K^\pm_{\mu3})/\Gamma(K^\pm_{e3})$. 
                         The $K_L$ measurements are corrected for comparison with $K^\pm$.}
      \label{fig:kmu3ke3}
    \end{minipage}
  \end{center}
\end{table}

More information is used when performing a global fit to all available kaon data. Such a fit, 
using more than 50 input measurements, has been done by the Flavianet working group~\cite{bib:palutan}.
The results on the parameter $r_{\mu e} = R_{K_{\mu3}/K_{e3}}^\text{Exp}/R_{K_{\mu3}/K_{e3}}^\text{SM}$ are
\be
r_{\mu e}^\pm   = 1.0059(87) \; \text{for} \; K^\pm, \quad \text{and} \quad 
r_{\mu e}^{L,S} = 1.0039(56) \; \text{for} \; K^0.
\ee
In the case of $K^\pm_{l3}$, the error is dominated by the experimental uncertainty.
For $K^0_{l3}$, both experimental and theoretical ($\delta_{EM}$) uncertainties contribute in similar magnitude.
Combining both results yields
\be
r_{\mu e}^{L,S} = 1.0042(50),
\ee
in excellent agreement with lepton universality. 
Furthermore, with a precision of $0.5\%$ the test in $K_{l3}$ decays 
has now reached the sensitivity of $\tau$ decays.

Not all the data of the experiments have been analyzed yet,
some improvement of the experimental accuracy can therefore be expected in the future.
In particular, improvements in the form factor measurements will reduce the experimental 
uncertainty further.
A significant reduction of the theoretical uncertainty, however, would require extensive two-loop calculations
for the radiative corrections.
At the end of the day, it is therefore likely that theory will 
limit lepton universality tests in $K_{l3}$ decays.

\begin{figure}[thb]
  \begin{center}
    \includegraphics[width=0.38\linewidth]{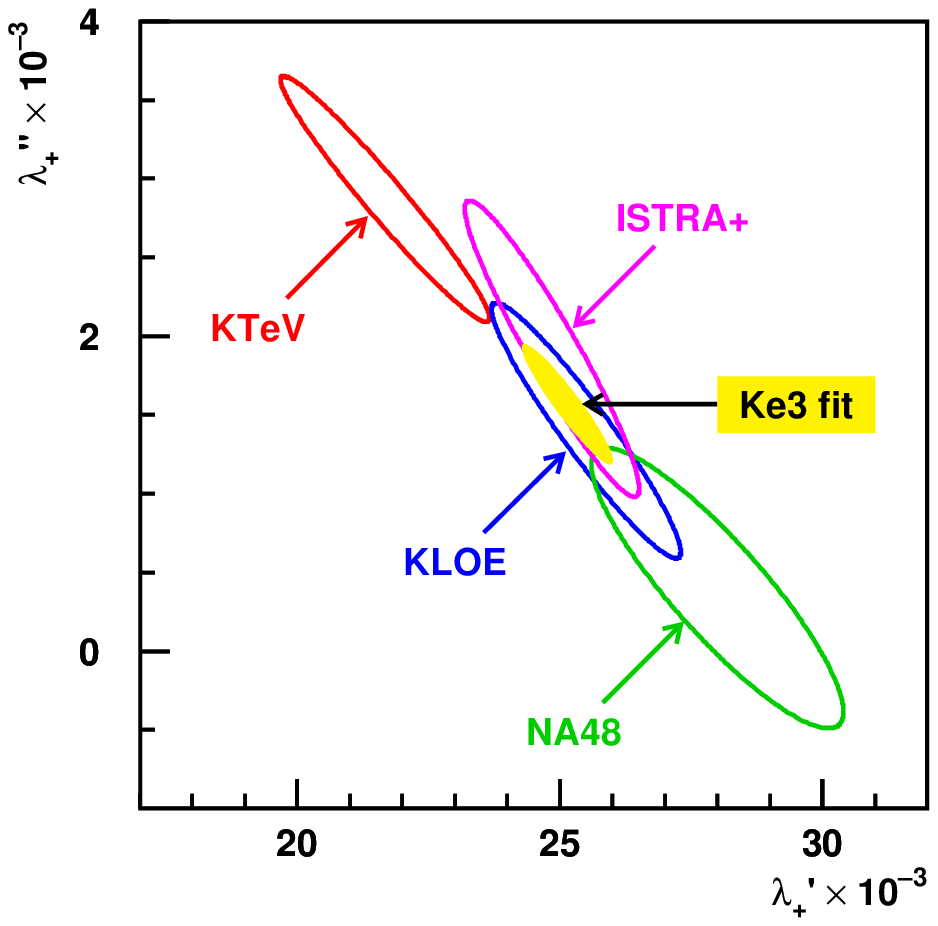}
    \hspace*{7mm}
    \includegraphics[width=0.33\linewidth]{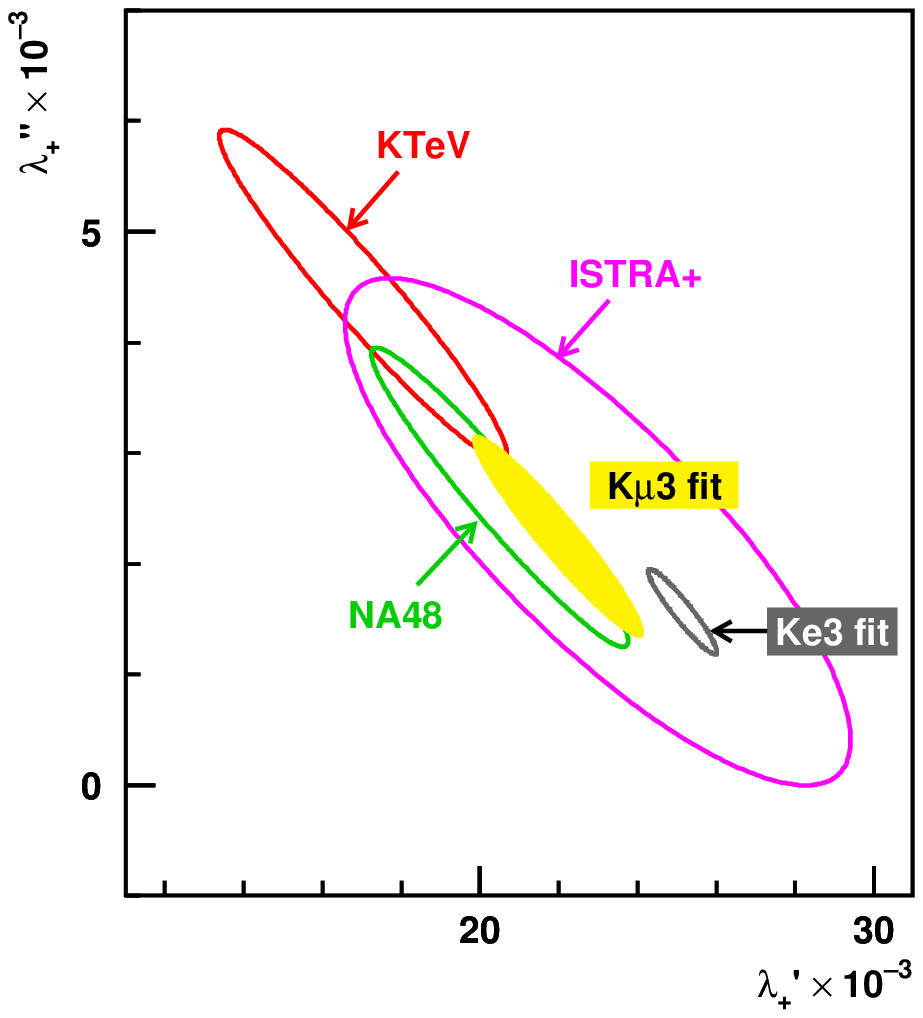}
    \caption{Measurements of the linear and quadratic slope parameter 
             of the vector form factor for $K_{e3}$ (left) and $K_{\mu3}$ (right).
             The ellipses are 68\% CL contours.}
    \label{fig:Kl3formfactors}
  \end{center}
\end{figure}

Another possibility to test lepton universality in $K_{l3}$ decays is the 
comparison of the vector form factor of $K_{e3}$ and $K_{\mu3}$.
For the $q^2$ dependency of the form factors several physically motivated parameterizations exist.
Recent experiments have commonly used the expansions
$f_+(q^2) =  f_+(0) \left( 1 + \lambda'_+ \, q^2/m_{\pi^+}^2 + \frac{1}{2} \lambda''_+ \, q^4/m_{\pi^+}^4 \right)$ 
for the vector form factor and
$f_0(q^2) = f_+(0) \left( 1 + \lambda_0 \, q^2/m_{\pi^+}^2 \right)$
for the scalar form factor.
The experimental results on the slope parameters are given in Tab.~\ref{tab:Kl3formfactors}
and are visualized --- including correlations between the parameters --- in Fig.~\ref{fig:Kl3formfactors}.
Also shown are the fits to the $K_{e3}$ and $K_{\mu3}$ data, respectively.
There is a mild disagreement between $K_{e3}$ and $K_{\mu3}$ in the vector form factor.
However, this is mainly driven by a single experiment: the NA48 result on $\lambda_0$ affects through correlations also 
$\lambda'_+$ and $\lambda''_+$. 
\begin{table}[t]
  \begin{center}
    \begin{tabular}{lcccc}
      \hline \hline
                                          & Channel   & $\lambda'_+ \times 10^3$ & $\lambda''_+ \times 10^3$ & $\lambda_0 \times 10^3$ \\ \hline 
      KTeV 2004~\cite{bib:KTeV_FF}        & $K_Le3$   & $21.7 \pm 2.0$           & $2.9 \pm 0.8$             &                         \\  
                                          & $K_L\mu3$ & $17.0 \pm 3.7$           & $4.4 \pm 1.5$             & $12.8 \pm 1.8$          \\ \hline
      KLOE 2006~\cite{bib:KLOE_Ke3FF}     & $K_Le3$   & $25.5 \pm 1.8$           & $1.4 \pm 0.8$             &                         \\  
      KLOE prel.~\cite{bib:KLOE_Kmu3FF}   & $K_L\mu3$ &  with $K_Le3$            &  with $K_Le3$             & $15.6 \pm 2.6$          \\ \hline
      NA48 2004~\cite{bib:NA48_Ke3FF}     & $K_Le3$   & $28.0 \pm 2.4$           & $0.4 \pm 0.9$             &                         \\  
      NA48 2007~\cite{bib:NA48_Kmu3FF}    & $K_L\mu3$ & $20.5 \pm 2.2$           & $2.6 \pm 0.9$             & $ 9.5 \pm 1.1$          \\ \hline 
      ISTRA+ 2004~\cite{bib:ISTRA_Ke3FF}  & $K^-e3$   & $24.9 \pm 1.7$           & $1.9 \pm 0.9$             &                         \\  
      ISTRA+ 2004~\cite{bib:ISTRA_Kmu3FF} & $K^-\mu3$ & $23.0 \pm 6.4$           & $2.3 \pm 2.3$             & $17.1 \pm 2.2$          \\
      \hline \hline
    \end{tabular}
    \caption{Measurements of $K_{l3}$ form factors.}
    \label{tab:Kl3formfactors}
  \end{center}
\end{table}

\section{Search for New Physics in $K_{\mu3}$ Decays}

The general Lagrangian of a charged current interaction can be written as
\be
{\cal L}_{CC} = \tilde{g} \left[ l_\mu + \frac{1}{2} 
                          \left( \begin{array}{c} {\scriptstyle u} \\*[-1mm] {\scriptstyle c} \\*[-1mm] {\scriptstyle t} \end{array} \right)
                          \left( {\cal V}_\text{eff} \gamma_\mu + {\cal A}_\text{eff} \gamma_\mu \gamma_5 \right)
                          \left( \begin{array}{c} {\scriptstyle d} \\*[-1mm] {\scriptstyle s} \\*[-1mm] {\scriptstyle b} \end{array} \right)
                          \right] W^\mu + \text{h.c.}.
\ee
In case of a purely left-handed SM-like interaction, the mixing matrices ${\cal V}_\text{eff}$ and ${\cal A}_\text{eff}$ are given by
${\cal V}_\text{eff} = - {\cal A}_\text{eff} = V_\text{CKM}$.
While the absence of right-handed currents in the weak interaction is well tested in 
the lepton sector, such tests are not available for charged currents in quark interactions.
As it has recently been pointed out~\cite{bib:Stern}, right-handed currents may be tested 
in the Dalitz plot distribution of $K_{\mu3}$ decays:
In the chiral limit,
the normalized scalar form factor $f_0(q^2)/f_+(0)$ is precisely known
at the Callan-Treiman point $q^2 = \Delta_{K\pi} = m_K^2 -m_\pi^2$ from 
branching fraction measurements. Right-handed currents, however, would cause a deviation from this prediction.
The crucial parameter, $\ln C \equiv \ln f(\Delta_{K\pi})$, can best be measured using a 
dispersive approach for the form factor parameterization.
The NA48 collaboration measures $\ln C = 0.1438 \pm 0.0080 \pm 0.0112$ in $K_{L\,\mu3}$ decays~\cite{bib:NA48_Kmu3FF},
which leads to an estimation of the parameters $\epsilon_S$, $\epsilon_{NS}$ of right-handed currents of
$2(\epsilon_S - \epsilon_{NS}) + \tilde{\Delta}_{CT} = -0.071 \pm 0.015$. Here,
$\tilde{\Delta}_{CT}$ is a $\chi$PT correction, which should be of ${\cal O}(10^{-3})$.
If confirmed by other experiments, this measurement would indicate right-handed currents in the quark sector.
However, the NA48 measurement of the form factor slope $\lambda_0$ is in disagreement with other 
experimental measurements. To what extent this affects also a disagreement in $\ln C$ needs to be seen in the future.

\section{Precision Measurements of $\Gemu$}

The ratio $R_K = \Gemu$ can be precisely calculated within the Standard Model. Neglecting radiative
corrections, it is given by 
\be
R_K^{(0)} = \frac{m_e^2}{m_\mu^2} \: \frac{(m_K^2 - m_e^2)^2}{(m_K^2 - m_\mu^2)^2} = 2.569 \times 10^{-5},
\ee
and reflects the strong helicity suppression of the electronic channel.
Radiative corrections have been computed within the model of vector meson dominance~\cite{bib:finkemeier},
yielding a corrected ratio of
\be
R_K = R_K^{(0)} ( 1 + \delta R_K^\text{rad.corr.} ) = 2.569 \times 10^{-5} \times ( 0.9622 \pm 0.0004 ) =(2.472 \pm 0.001) \times 10^{-5}.
\ee

Because of the helicity suppression of $K_{e2}$ in the SM, the decay amplitude is a prominent candidate
for possible sizeable contributions from new physics beyond the SM. Moreover, when normalizing to $K_{\mu2}$ decays,
it is one of the few kaon decays for which the SM-rate is predicted with very high accuracy.
Any significant experimental deviation from the prediction would immediately be evidence for new physics.
However, this new physics would need to violate lepton universality to be visible in the ratio $K_{e2}/K_{\mu2}$.

Recently it has been pointed out, that in a SUSY framework sizeable violations of lepton universality can be expected
in $K_{l2}$ decays~\cite{bib:masiero}. At tree level, 
lepton flavour violating terms are forbidden in the MSSM. 
Loop diagrams, however, should induce lepton flavour violating Yukawa couplings as  $H^+ \to l \nu_\tau$ 
to the charged Higgs boson $H^+$.
Making use of this Yukawa coupling, the dominant non-SM contribution to $R_K$ modify the ratio to
\be
R_K^\text{LFV} \approx R_K^\text{SM} \left[ 1 + \left( \frac{m_K^4}{M_{H^\pm}^4} \right) \left( \frac{m_\tau^2}{M_e^2} \right) |\Delta_{13}|^2 \tan^6 \beta \right].
\label{eqn:susy}
\ee
The lepton flavour violating term $\Delta_{13}$ should be of the order of $10^{-4}-10^{-3}$, as expected from 
neutrino mixing. F
or moderately large $\tan \beta$ and $M_{H^\pm}$, SUSY contributions may therefore enhance $R_K$ by up to a few percent.
Since the additional term in Eqn.~\ref{eqn:susy} goes with the forth power of the meson mass, no similar effect
is expected in $\pi_{l2}$ decays.

Experimental knowledge of $K_{e2}/K_{\mu2}$ has been poor so far. The current world average
of $R_K = (2.45 \pm 0.11) \times 10^{-5}$ dates back to three experiments of the 1970s~\cite{bib:pdg}
and has a precision of less than 4\%.
However, now three new preliminary measurements were reported by NA48/2 and KLOE (see Tab.~\ref{tab:ke2kmu2}).
A preliminary result of NA48/2, based on about 4000 $K_{e2}$ events from the 2003 data set, 
was presented in 2005~\cite{bib:Ke2_2003}.
Another preliminary result, based on also about 4000 events, recorded in a minimum bias run period in 2004, 
was shown at KAON07\cite{bib:Ke2_2004}.
Both results have independent statistics and are also independent in the systematic uncertainties, 
as the systematics are either of statistical nature (as e.g.\ trigger efficiencies) or determined in
an independent way. The somewhat larger systematic uncertainty of the result on the 2004 data is explained
by a different method of background rejection, which relies only on data statistics.
Another preliminary result, based on about 8000 $K_{e2}$ events, was presented at KAON07
by the KLOE collaboration~\cite{bib:Ke2_KLOE}.
Both, KLOE and NA48/2 measure the inclusive ratio $\Gamma(K_{e2(\gamma)})/\Gamma(K_{\mu2(\gamma)})$.
The small contribution of $K_{l2\gamma}$ events from direct photon emission from the decay vertex 
was subtracted by each of the experiments.
Combining these new results with the current PDG value yields a current world average of
\be
R_K = \Gemu = ( 2.457 \pm 0.032 ) \times 10^{-5},
\label{eqn:ke2kmu2}
\ee
in very good agreement with the SM expectation and, with a relative error of $1.3\%$,
a factor three more precise than the previous world average (Fig.~\ref{fig:ke2kmu2}).

\begin{table}[t]
  \begin{center}
    \begin{minipage}[t]{0.54\linewidth}
      \begin{tabular}{lc}
        \hline \hline
                                                  & $\Gemu$ $[10^{-5}]$  \\ \hline
        PDG 2006~\cite{bib:pdg}                   & $2.45 \pm 0.11$ \\
        NA48/2 prel.\ ('03)~\cite{bib:Ke2_2003}   & $2.416 \pm 0.043 \pm 0.024$ \\
        NA48/2 prel.\ ('04)~\cite{bib:Ke2_2004}   & $2.455 \pm 0.045 \pm 0.041$ \\
        KLOE prel.~\cite{bib:Ke2_KLOE}            & $2.55 \pm 0.05 \pm 0.05$ \\ \hline
        SM prediction                             & $2.472 \pm 0.001$ \\
        \hline \hline
                                                  & \\*[-3mm]
      \end{tabular}
      \caption{Results and prediction for $R_K = \Gemu$.}
      \label{tab:ke2kmu2}
    \end{minipage}
    \hspace*{\fill}
    \begin{minipage}[t]{0.43\linewidth}
      \raisebox{-17mm}{\includegraphics[width=\linewidth]{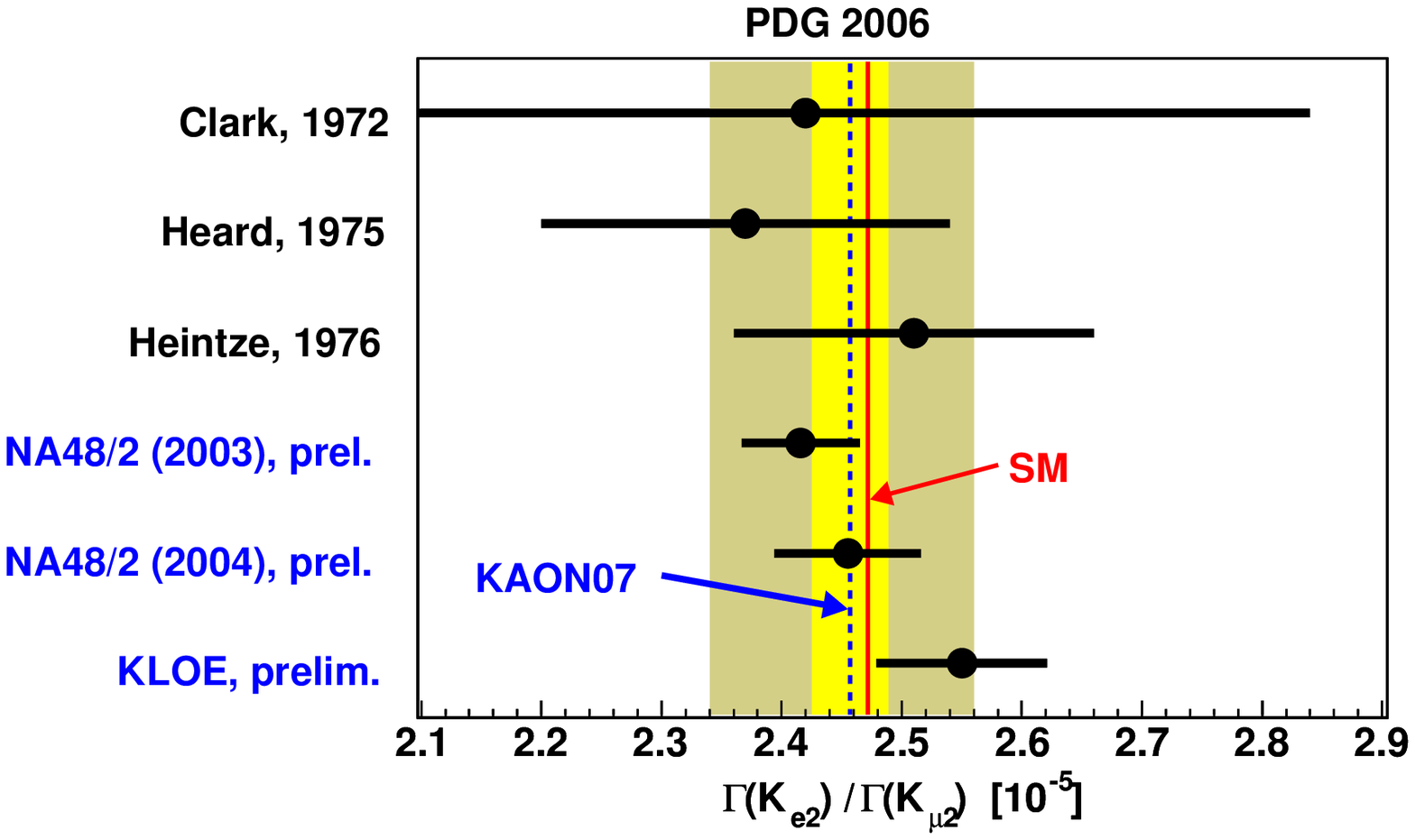}}
      \captionof{figure}{Results on $R_K = \Gemu$. Indicated are the current PDG average (brown),
                        the average of all measurements (blue/yellow), and the SM prediction (red).}
      \label{fig:ke2kmu2}
    \end{minipage}
  \end{center}
\end{table}

\begin{figure}[t]
  \begin{center}
    \includegraphics[width=0.49\linewidth]{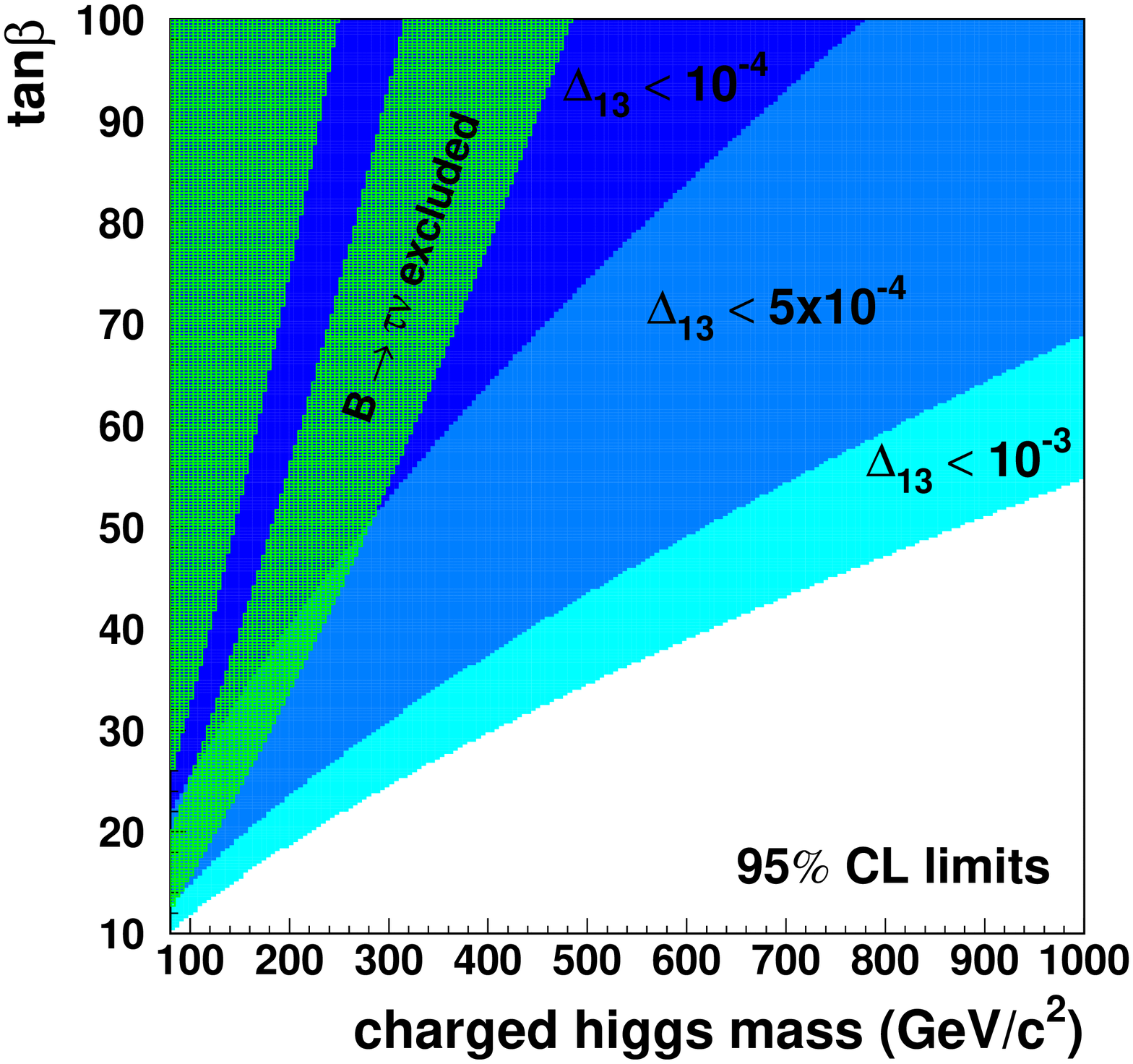}
    \hspace*{\fill}
    \includegraphics[width=0.49\linewidth]{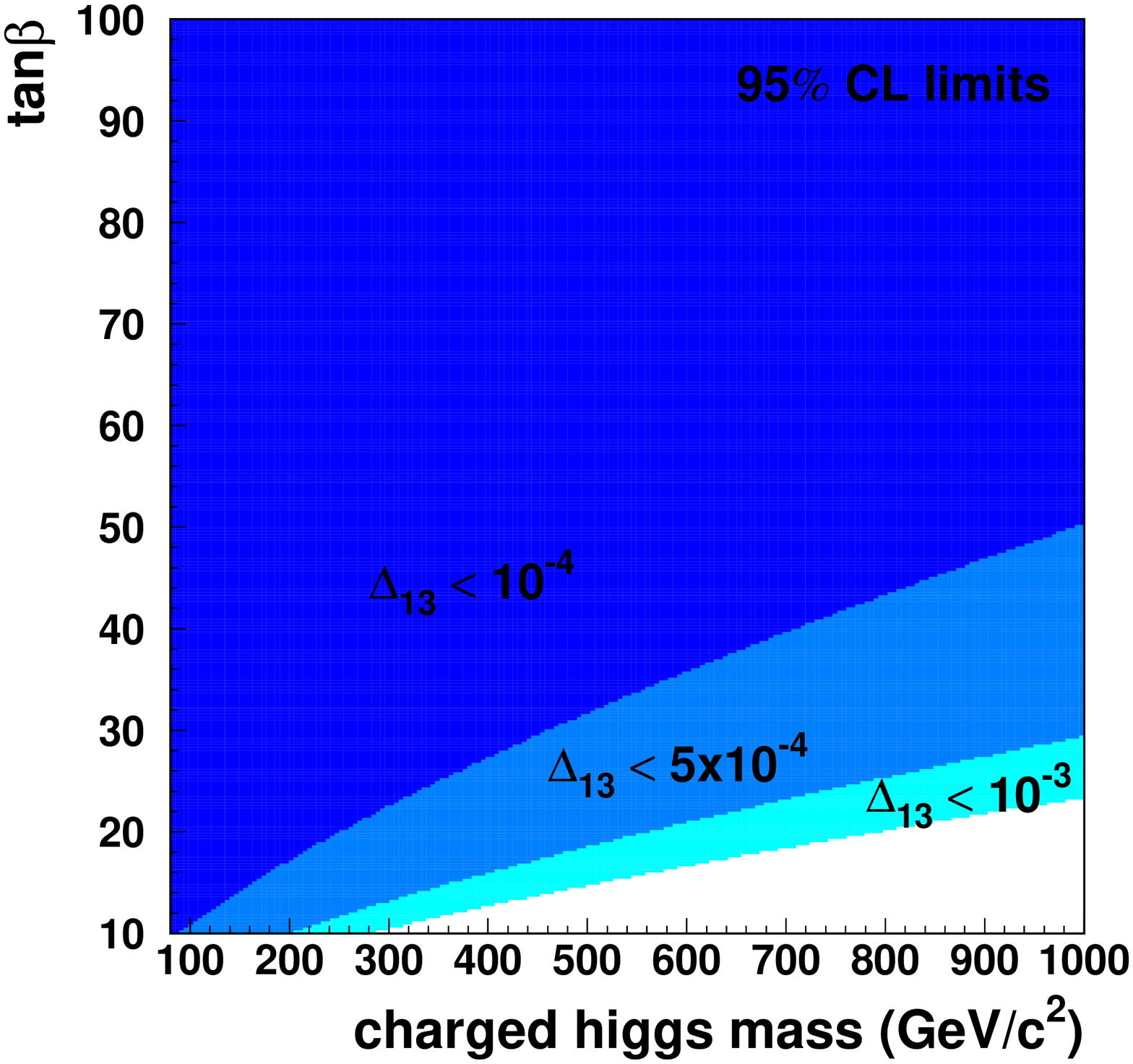}
    \caption{Exclusion limits at $95\%$ CL on $\tan \beta$ and the charged Higgs mass $M_{H^\pm}$ from $\Remu$ for different values of $\Delta_{13}$. 
             Values $M_{H^\pm} < 80$~GeV/$c^2$ are excluded by direct searches. 
             {\em Left:} limits from the current world average on $R_K$. Indicated are also the 
             exclusion limits from $B \to \tau \nu_\tau$ decays~\cite{bib:btaunu}.
             {\em Right:} prospects of exclusion limits for the same mean $R_K$, but with a relative uncertainty of $0.3\%$,
             as expected from the upcoming P326 measurement.}
    \label{fig:susylimit}
  \end{center}
\end{figure}

In the SUSY framework discussed above, this result gives strong constraints for $\tan \beta$ and $M_{H^\pm}$ (Fig.~\ref{fig:susylimit} (left)).
For a moderate value of $\Delta_{13} \approx 5 \times 10^{-4}$, $\tan \beta > 50$
is excluded for charged Higgs masses up to 1000~GeV/$c^2$ at 95\% CL.
These exclusion limits can be compared with SUSY limits obtained from $B \to \tau \nu_\tau$ decays. The standard model rate
of this decay is modified by tree-level charged Higgs exchange to~\cite{bib:isidori} 
\be
\Br(B \to \tau \nu_\tau)_\text{SUSY} \; = \; \Br(B \to \tau \nu_\tau)_\text{SM} \: 
                                             \left[ 1 - \left( \frac{m_B^2}{M_{H^\pm}^2} \right) \frac{\tan^2 \beta}{1 + \epsilon_0 {\tan \beta}} \right]^2,
\label{eqn:btaunu}
\ee
with $\epsilon_0$ of the order of 0.01.
The exclusion limits on $\tan \beta$ and $M_{H^\pm}$, obtained from
the current BELLE and BaBar average of $\Br(B \to \tau \nu_\tau) = (1.42 \pm 0.44) \times 10^{-4}$~\cite{bib:btaunu},
are superposed in Fig.~\ref{fig:susylimit} (left). 
In general, the limit obtained from $\Remu$ is stronger than those from $B \to \tau \nu_\tau$. However,
as the latter are lepton flavour conserving, they do not need the assumption on the value $\Delta_{13}$.

In the near future, further improvements in the knowledge of $R_K$ are expected.
The preliminary KLOE measurement~\cite{bib:Ke2_KLOE} 
is statistically dominated by MC statistics and has a conservative estimate of the systematic uncertainty.
Improving both these items, adding the remaining KLOE statistics, and also adding events with an 
additional reconstruction method, should reduce the overall relative uncertainty down to 1\%.

\begin{figure}[t]
  \begin{center}
    \includegraphics[width=0.54\linewidth]{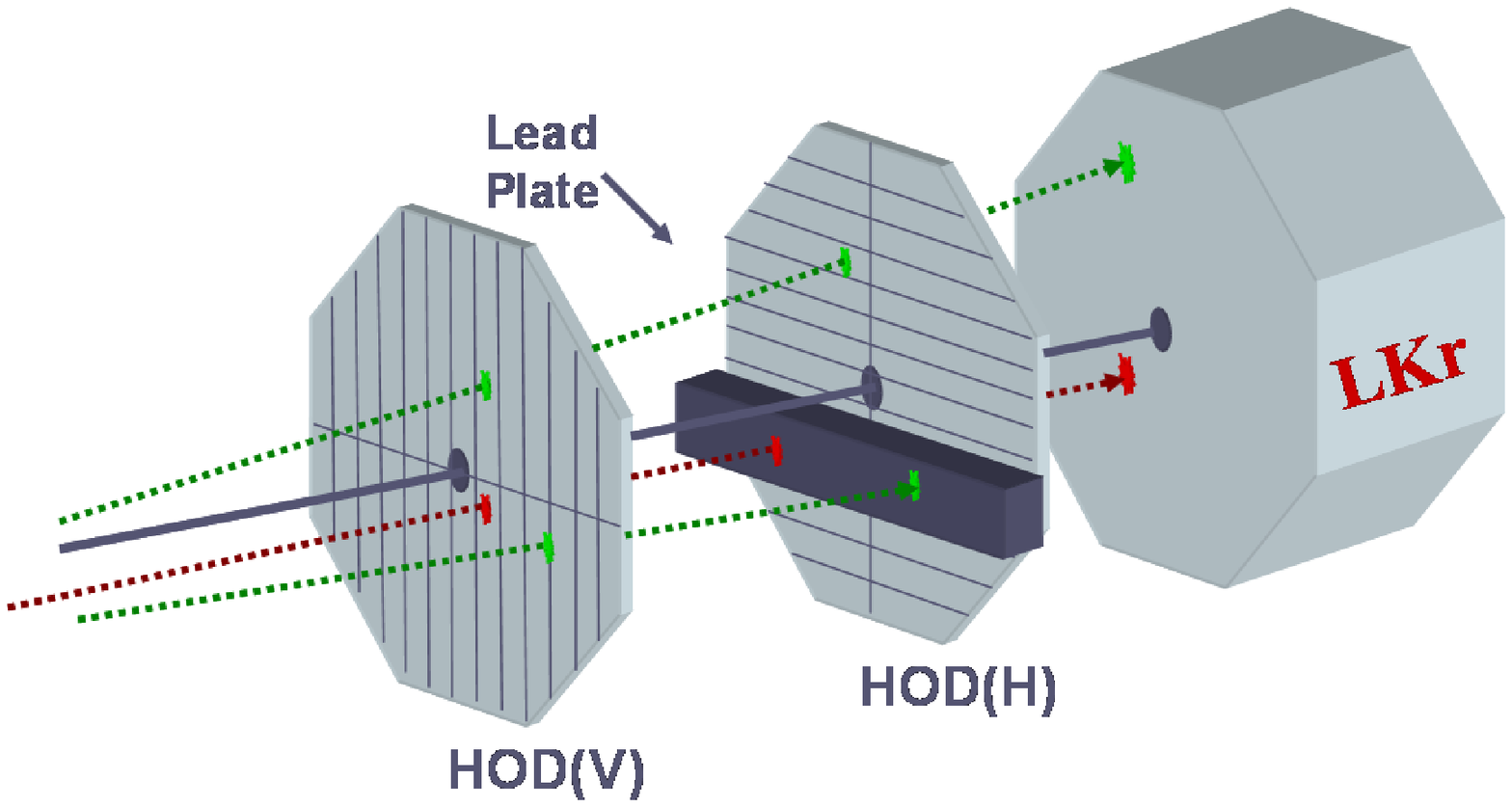}
    \hspace*{\fill}
    \includegraphics[width=0.42\linewidth]{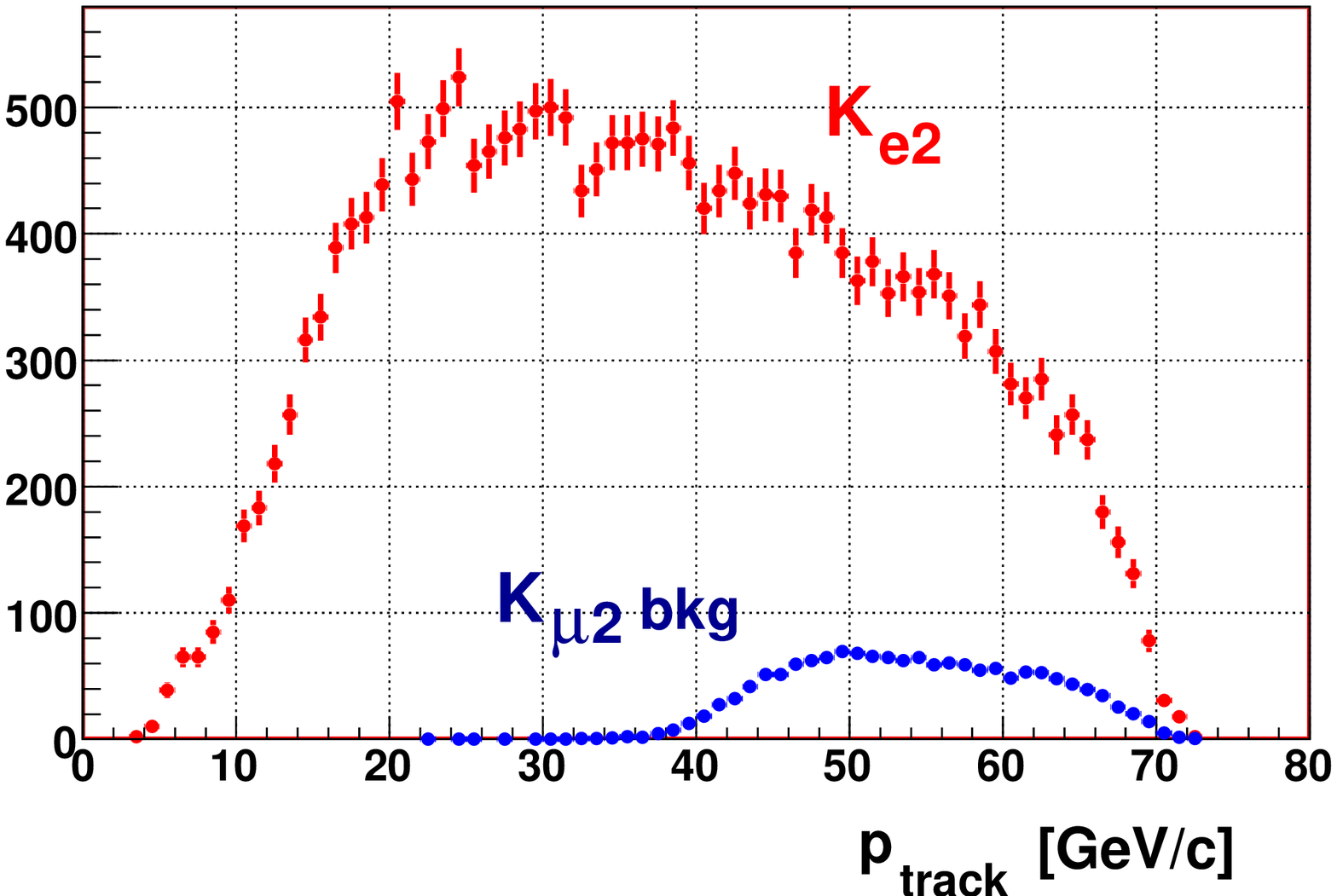}
    \caption{Background from $K_{\mu2}$ to $K_{e2}$ in the P326 run 2007.
             {\em Left:} lay-out of the lead bar before the calorimeter front face.
             {\em Right:} $K_{\mu2}$ background as function of the lepton track momentum.}
    \label{fig:P326bkg}
  \end{center}
\end{figure}

An improvement of a further factor of three is the goal of the P326 collaboration at CERN, 
which in 2007 performs a dedicated run period to measure $R_K$.
The P326 experiment, successor of NA48, plans to measure the very rare decay $\kppinunu$ in the mid-term future~\cite{bib:P326}.
In a first phase, about four months of dedicated data taking take place this year,
with the aim of recording 150000 $K_{e2}$ decays in total~\cite{bib:Ke2_P326}.
For this measurement, beamline and set-up are very similar to the NA48/2 experiment, with some minor adjustments to
optimize the signal acceptance and the kinematic separation from the background. 
To select one-track events, a minimum bias trigger with an efficiency larger than 99\% is employed, 
requiring the hit pattern in the scintillator hodoscope to be compatible with at least one charged track.
In addition, a minimum energy deposit in the electromagnetic calorimeter is required 
to suppress and downscale the dominant $K_{\mu2}$ decay.
Since all other kaon decays and non-kaon events can easily be rejected kinematically,
the major source of background to $K_{e2}$ are $K_{\mu2}$ events. Also for these events kinematic suppression is sufficient, as
long as the track momentum does not exceed 35~GeV/$c$ (43\% of all events). 
For events above this threshold, electrons have in addition to be identified by the
ratio $E/p$ of energy deposited in the calorimeter and momentum measured in the spectrometer.
The small fraction of about $5\times 10^{-6}$ of $K_{\mu2}$ events for which the muon loses
all its energy in the calorimeter by catastrophic bremsstrahlung is determined from data:
A lead bar, covering about 18\% of the calorimeter acceptance (Fig.~\ref{fig:P326bkg}), allows
to select purely muonic events, thus permitting to measure the $E/p$ distribution of muons 
from data.
The uncertainty on the background determination is purely statistical and will be of order 0.1\%.
All other sources of systematic uncertainties as trigger efficiencies, radiative corrections, and electron identifications
are mostly of statistical nature, too, and do not exceed $\pm 0.2$. This has been demonstrated
in the NA48/2 measurement on the 2004 data set~\cite{bib:Ke2_2004}, which was performed under very similar conditions, 
but is still limited by the statistics of the background suppression.
The total uncertainty of the P326 measurement on $R_K$ is expected to be $\pm 0.34\%$.
This would improve the current world average by another factor of four,
thus allowing to either find a deviation from the SM expectation 
or to set very stringent limits in the SUSY parameter space (Fig.~\ref{fig:susylimit} (right)).

\section{Conclusion}

Large improvements in lepton universality tests have been achieved in kaon physics 
in the recent few years.
In $K_{l3}$ decays, the ratio $\Rmue$ is now in agreement with the SM expectation on a level of $0.5\%$.
For further improvement, theoretical uncertainties now start to be important.
For the ratio $\Gemu$, two new preliminary results have been quoted by KLOE and NA48/2.
Together with older measurements, this ratio has now a precision of $1.3\%$, which is an improvement
of a factor of three w.r.t.\ 2004. It is in perfect agreement with the SM prediction, which can be turned into
strong constraints in the SUSY parameter space of $\tan \beta$ and $M_{H^\pm}$.
Further improvements on $\Gemu$ are expected by the final KLOE result and a dedicated run of the NA48 successor P326
in 2007, aiming at a precision of a few per mil.

\end{document}

%% file: pretext.tex
\def\Journal#1#2#3#4{{#1}~{\bf #2} (#4), #3}

\def\PLB{{\em Phys.~Lett.}~B}
\def\PRL{\em Phys.~Rev.~Lett.}
\def\PRD{{\em Phys.~Rev.}~D}
\def\ZPC{{\em Z.~Phys.}~C}
\def\EPJ{{\em Eur.~Phys.~J.}~C}
\def\JPG{{\em J.~Phys.}~G}

\newcommand{\Br}{{\text{Br}}}

\newcommand{\Remu}{K_{e2}/K_{\mu2}}
\newcommand{\Gemu}{\Gamma(K_{e2})/\Gamma(K_{\mu2})}
\newcommand{\Rmue}{K_{\mu3}/K_{e3}}
\newcommand{\Gmue}{\Gamma(K_{\mu3})/\Gamma(K_{e3})}

\newcommand{\kppinunu}{K^+ \to \pi^+ \nu \bar{\nu}}

\newcommand{\bdm}{\begin{displaymath}}
\newcommand{\edm}{\end{displaymath}}
\newcommand{\be}{\begin{equation}}
\newcommand{\ee}{\end{equation}}
